\documentclass[12pt]{article}
\usepackage{graphicx}
\usepackage{amssymb}

\def\bq{\begin{equation}}
\def\eq{\end{equation}}
\def\ba{\begin{array}}
\def\ea{\end{array}}
\def\e{\varepsilon}
\def\sech{\,\mbox{sech}}

\begin{document}

{\huge   A new criterion for the existence of KdV solitons in
ferromagnets.}
\begin{center}
 {\Large  H.Leblond  }
  \vspace{1cm}\\
Laboratoire P.O.M.A., U.M.R.-C.N.R.S. 6136\\
Universit\'e d'Angers\\
2 $\rm B^d$ Lavoisier 49045 ANGERS Cedex 01, France.\\
\vspace{2cm}
 {\bf Abstract}
\end{center}
\noindent

The long-time evolution of the KdV-type  solitons propagating  in ferromagnetic materials is
considered trough a multi-time formalism,
it is governed by
all equations of the KdV Hierarchy.
The scaling coefficients of  the higher order
time variables  are explicitly computed in terms of the physical parameters,
showing that the
KdV asymptotic is valid only when the angle between the
propagation direction and the external magnetic  field is  large enough.
The one-soliton solution of the KdV hierarchy is written down in terms of the physical parameters.
A maximum value of the soliton parameter is determined, above which the perturbative approach is not valid.
Below this value, the  KdV soliton conserves its properties during an infinite propagation time.

\vspace{1cm}

P.A.C.S. :  03.40.K (41.20.J, 75.50.G).

\newpage
\section{Introduction}
\subsection{KdV-type solitons in ferromagnets}

Electromagnetic wave propagation in ferromagnetic media is intrinsically
nonlinear. It is therefore the matter of intensive research in the theoretical physics
of the nonlinear waves. In the frame of the Maxwell-Landau model, analytical
expressions describing solitary wave propagation out from any slowly envelope or long-wave
approximation have been found \cite{bas88}. These waves have also been studied  numerically
\cite{ostrov}. Envelope solitons have been studied from several theoretical approaches
\cite{bas90a,sla94a}. There are many experiments  regarding magnetostatic waves in thin films \cite{deg87,cottam,bau98a}.
Long-wave type approximations allow to describe some features related to relativistic domain wall
propagation \cite{nak91c, mkp}, but have also brought forward the existence of another  type of wave, described by
the Korteweg-de Vries (KdV) equation \cite{twowa}. It has been shown that such a wave can be emitted by a transverse
instability of the relativistic domain wall \cite{ostrov,mkp}.
The interaction between the two types of  wave has also been studied \cite{twowa}.

The KdV model is obviously a rough approximation. In \cite{twowa}
where it has been first derived in this frame, anisotropy, damping and inhomogeneous exchange
were neglected. Second it assumes that the  wave depends on a single spatial coordinate (plane wave), and
that the amplitude is
weak enough, the wave length and the propagation distance large  enough, so that the
first order of the KdV approximation can be retained.
A study taking into account three space dimensions, damping, and inhomogeneous exchange is published independently
\cite{kpburg}.
But the weakly nonlinear approximation itself may necessitate higher order corrections.
The latter are independent from the former ones.
Indeed, the wave is intrinsically nonlinear, and the weakly nonlinear approximation
is forced by the introduction of a static field, to which the wave field can be compared.
Even in the roughest approximation, the ratio between the two fields can become rather close to one.
A derivation of the equations describing the evolution of the higher order terms
has been derived using a multi-time formalism \cite{hierarchy}. It allows to prove that a formal asymptotic expansion
exists up to any order, with all its terms bounded \cite{linkdv}, which is a first step in the mathematical justification
of the convergence of the expansion.
Following the idea by Kraenkel {\it et al.} \cite{krapb,manna3}, the multi-times expansion for KdV uses the KdV Hierarchy.
The evolution of the main term in the expansion relative to each higher order time variable
is given by the corresponding equation in the KdV Hierarchy. Regarding the main term only, all information about the
particular physical situation considered is contained in scaling coefficients of the time variables.
These coefficients can be computed. 
The aim of this paper is to give the value of these
quantities, and to draw physical consequences from them.
It is organized as follows: in section 2 we describe the perturbative scheme in the multi-time formalism.
In section 3 we compute explicitly the time scaling coefficients. Conclusions can be drawn on the validity
of the pertubative scheme considered as an asymptotic expansion,  {\it i.e.} for a fixed number of terms,
when the perturbative parameter becomes small enough. In section 4, we give the expression of the
one-soliton solution of the complete KdV hierarchy. This gives information about
the validity of the perturbative scheme considered as a series expansion, {\it i.e.} for a fixed value of
the perturbative parameter and an infinite number of terms.

\section{The multi-time formalism}
\subsection{The KdV mode}

The evolution of  the
magnetization density $\vec M$ in a magnetic field
$\vec H$ is described by the Landau equation
\begin{equation}
\partial _{t}\vec{M}=-\gamma\mu_0\vec{M}\wedge \vec H_{eff} , \label{ferri1}
\end{equation}
where $\gamma$ is the  gyromagnetic  ratio ($\gamma>0$), and
$\mu_0$ the magnetic permeability in vacuum.
The effective
field $\vec H_{eff}$ contains several  terms  giving account for
the inhomogeneous exchange interaction, the effects of finite size  and  the anisotropy.
Here we use the basic approximation: $\vec H_{eff}=\vec H$.
Damping is also neglected.

The evolution of the magnetic field $\vec H$ is described by the
Maxwell equations. We assume that, regarding  its  dielectric
properties, the material is perfectly linear and isotropic, and we
denote by $c$ the light velocity based on its dielectric constant
$\hat\e$, {\it i.e.} $c=1/\sqrt{\hat\e\mu_0}$. The Maxwell
equations reduce then to \bq -\vec\nabla\left(\vec\nabla\cdot\vec
H\right)+\Delta\vec H=\frac1{c^2}\partial_t^2\left(\vec H+\vec
M\right). \label{ferrimaxwell}\eq
 We  replace below
$\vec H$, $ \vec M$ and $t$ by the  normalized quantities
$\gamma\mu_0\vec H/c$, $\gamma\mu_0\vec M/c$ and $ct$.  The
constants $\gamma\mu_0$ and $c$ take then the value 1.

 The  `long wave'
limit of a wave with negative helicity is considered.
We introduce a small parameter $\e$, such that $1/\e$ measures the length of the solitary wave
and $\e^2$ its amplitude.
The magnetic field is expanded as
\bq\vec H=\vec  H_0+\e^2 \vec H_2+ \cdots,\eq
and $\vec  M$ in an analogous way.
Using the slow variables
\bq\left\{\ba{ll}\xi=&\e(x-Vt),\\
\tau_1=&\e^3t,\ea\right.\eq
it is shown first that this wave propagates at the velocity
\bq V=\sqrt{\left(\alpha+\sin^2\theta\right)/\left(\alpha+1\right)},\label{vit}\eq
where $\theta$ is the angle between the propagation direction and
the  applied field, and $\alpha=H_0/M_0$ the ratio from the latter to the
saturation magnetization. Second it is shown that the propagation of this type of
`long waves' is governed by the KdV equation~\cite{twowa} \bq
\partial_{\tau_1}\varphi_{2}+q \varphi_2\partial_\xi\varphi_{2}+r\partial_\xi^3 \varphi_{2}=0,\label{ferrikdv}\eq
where $q$ and $r$ are real constants given by
\bq
q=\frac32\frac{\cos^2\theta\sin^2\theta\sqrt{1+\alpha}}
{\left(\alpha+\sin^2\theta\right)^{3/2}},\eq
and
\bq
r=\frac{-1}{2m^2}\frac{\cos^4\theta\sqrt{\alpha+\sin^2\theta}}
{\sin^2\theta\left(1+\alpha\right)^{7/2}}.\eq
 $\varphi_2$ is
the wave amplitude, related to the main  component $\vec H_2$ and $\vec M_2$ of the wave magnetic field
and the magnetization density through
\bq
\vec H_2=\varphi_2\vec h_1\qquad \mbox{and}\qquad \vec M_2=\varphi_2\vec m_1,\eq
where $\vec h_1$ and $\vec m_1$ are polarization vectors defined by
\bq \vec h_1=m\left(1+\alpha\right)\sin\theta\left(\ba{c}
\frac{\sin\theta\cos\theta}{\alpha+\sin^2\theta}\\1\\0\ea\right),\label{achun}\eq
and
\bq \vec m_1=\frac{m\left(1+\alpha\right)\sin\theta\cos\theta}
{\alpha+\sin^2\theta}
\left(\ba{c}-\sin\theta\\\cos\theta\\0\ea\right).\eq
(We use the normalization of \cite{twowa,hierarchy}, introduced for computational convenience).

\subsection{Higher order terms}
Going further in the resolution of the perturbative scheme, it is seen that the field component of order $j$
($j>2$)  writes
\bq
\vec H_j=\varphi_j\vec h_1+\vec h_j^0\left(\varphi_2,\varphi_3,\ldots,\varphi_{j-1}\right),\eq
where $\vec h_j^0$ is an explicit functional of the lower  order amplitudes $\varphi_2$, $\varphi_3$, up to $\varphi_{j-1}$,
$\vec h_1$ the polarization vector given above by (\ref{achun}), and
$\varphi_j$ is an higher order amplitude.
$\varphi_j$
 satisfies a linearized KdV equation of the form
\bq\ba{r}\displaystyle
\partial_{\tau_1}\varphi_j+q\partial_\xi(\varphi_2\varphi_j)
+r\partial_\xi^3\varphi_j=\hspace{2cm}\vspace{1mm}\\\displaystyle
\Xi_j(\varphi_2,\varphi_3,\ldots\varphi_{j-1}),\ea
\label{58}\eq
where the right-hand-side (rhs) member $\Xi_j$ is an explicit functional of the lower
 order amplitudes $\varphi_2$, $\varphi_3$, up to $\varphi_{j-1}$.
 The
  parity and  homogeneity properties of the expansion allow to prove that half of these equations
admit the zero solution, so that $\varphi_j$ is non zero for even $j$ only.
Note that the inhomogeneous part $\vec h_j^0$ of the $j^{\rm th}$ order magnetic field amplitude $\vec H_j$
does not vanish for odd $j$.

 We study the long-time
 propagation by considering the unbounded   or
secular solutions, and a  multi-time expansion. Therefore  we
introduce a sequence of slower and slower temporal variables
 $\tau_1=\tau$, $\tau_2$, $\tau_3$,... defined by
$\tau_j=\e^{2j+1}t$. The propagation is
 governed by all  equations of the KdV Hierarchy.
In particular, the equation  giving
the evolution of the leading term $\varphi_2$ with regard to the first higher order time variable
$\tau_2$ is derived as follows (more detail is given in \cite{hierarchy}).
$\varphi_4$ is the  amplitude  of the first correction to the main term whose amplitude
is $\varphi_2$.  The equation
that determines its evolution  can be written in
the form
 \bq
\partial_{\tau_1}\varphi_{4}+q\partial_\xi\left(\varphi_2\varphi_{4}\right)
+r\partial_\xi^3\varphi_{4}=-\partial_{\tau_2}\varphi_2-r_2\partial_\xi^5\varphi_2+{\cal
O}_2 \label{hier4},\eq
where ${\cal O}_2$ refers  to an expression
depending on $\varphi_2$ without  linear term, and $r_2$ is
a real coefficient.
Some functions $\varphi_4$ solutions of (\ref{hier4}) are secular, {\it i.e.}
grow linearly with the time $\tau_1$.
Consider values of the time variable $t$ about $1/\e^5$.
Then the the time variable $\tau_1=\e^3 t$ take values about  $1/\e^2$,
and the secular term in $\varphi_4$ becomes of order $\e^2$ instead of
$\e^4$, due to the factor  $\tau_1\propto 1/\e^2$.
For times with this order of magnitude,
this  correction term must be  taken  into account in the expression of
the main amplitude $\varphi_2$.
In order to
incorporate the correction  into the evolution of the main amplitude $\varphi_2$ with regard to the second order time
variable $\tau_2$,
we impose some condition on the rhs member of equation (\ref{hier4}), so that  $\varphi_4$  remains  bounded (or more exactly sublinear).
The
condition to be satisfied is thus that the equation (\ref{hier4})
does not admit any secular solution. Through an explicit
computation
 in the case where $\varphi_2$ is the one-soliton solution
of KdV, Kodama and Taniuti \cite{kod78t} have noticed that the
secular-producing terms are the terms linear with regard to the
solution of lowest order  $\varphi_2$. The secular solutions
$\varphi_4$  will vanish thus if the linear terms
 vanish from the rhs member of the equation (\ref{hier4}).
To achieve this, we impose that $\varphi_2$  satisfies some
partial differential equation such that
\bq\partial_{\tau_2}\varphi_2=-r_2\partial_\xi^5\varphi_2+{\cal
O}_2.\label{hier5}\eq
 We  still need to determine the nonlinear
terms
 of  equation (\ref{hier5}), represented by ${\cal O}_2$. They are not
free but imposed by the compatibility condition between the KdV
equation (\ref{ferrikdv}) and the equation (\ref{hier5}), which
is the Schwartz condition:
 $\partial_{\tau_1}\partial_{\tau_2}\varphi_2=\partial_{\tau_2}\partial_{\tau_1}\varphi_2$.
Kraenkel, Manna, and Pereira \cite{manna3} have conjectured and
checked on many examples that the only  equation that possesses
the same homogeneity  properties as the rhs member of
(\ref{hier4}), and that satisfies this condition, is the second
equation of what is called the KdV Hierarchy.

The  KdV Hierarchy is the  following family of equations
\cite{fnt}:
\bq\partial_{T_n}v=\partial_X {\cal L}^nv\hspace{2cm}
\mbox{($n$ integer)},\label{hier6}\eq
where $\cal L$ is a
recurrence operator, defined by
\bq{\cal L}=-\frac14\partial_X
^2-v+\frac12\int^X dX (\partial_X v).\label{hier7}\eq For  $n=1$,
it is the KdV equation, with a normalization that differs from
that of (\ref{ferrikdv}) ($q=\frac32$~, $r=\frac14$). We identify
both using the relations \bq
v=\frac{q}{6r}\varphi_2\hspace{5mm},\hspace{1cm}X=\xi
\hspace{5mm}\mbox{and}\hspace{1cm}
T_1=4r\tau_1\label{hier7b}.\eq For $n=2$, the equation
 of the  Hierarchy (\ref{hier6}) writes as \bq
\partial_{T_2}v=\frac1{16}\partial_X ^5v+\frac54(\partial_X
v)\partial_X ^2v + \frac58v\partial_X ^3v +
\frac{15}8v^2\partial_X v\label{hier8}.\eq An important property
 is the existence of the  $\tau$  Hirota function
\cite{fnt}, that is a function of all  variables
$(X,T_1,T_2,\ldots)$, related to $v$ by
$$v(X ,T_1,T_2,\ldots)=2\partial_X ^2\ln\tau(X ,T_1,T_2,\ldots)$$
(avoid any  confusion between the  $\tau$   Hirota function and
the time variables  $\tau_j$). The existence of $\tau$ ensures
that a solution $v$ of the system yielded by all equations of the
Hierarchy exists, and thus that
the Schwartz condition  is  satisfied at any  order. After an
adequate choice  of the proportionality constant that connects the
 time variables of order 2, the variable  $\tau_2$ of our expansion and
the variable $T_2$ of the Hierarchy, that we write as
 \bq T_2=-16r_2\tau_2,\eq
 the evolution equation  to be satisfied by
$\varphi_2$ is  \bq
\frac{-1}{16r_2}\partial_{\tau_2}\varphi_2=\partial_\xi{\cal
L}^2\varphi_2 .\label{hier9}\eq This way,  the linear terms  have
been removed from the equation (\ref{hier5}). It remains to
justify that this procedure, that removes all linear terms  from
the rhs member of the linearized KdV equation,
assuming it polynomial with regard to the solution of KdV, ensures
that the solution of the linearized equation is bounded
\cite{secular}.
The KdV equation admits an infinite sequence of conserved densities
we denote by  ${\cal A}_j$, an expression of which can be found in \cite{kod78t}.
It has been proven in \cite{secular}  that the secular-producing terms
are the terms proportional to $\partial_\xi{\cal A}_j$.
Further the relations  existing between the
conserved densities ${\cal A}_j$,  and the recurrence operator
 $\cal L$, defined by (\ref{hier7}), that allows to write the Hierarchy,
 allow to show that the procedure by Kraenkel {\it et al.},
 initially intended to remove the linear
terms,
exactly removes  all these secular-producing terms.

Otherwise, the rhs member of the linearized KdV
equation
 that governs the evolution of $\varphi_6$ involves $\varphi_4$, solution of (\ref{hier4}). It is thus
necessary to see wether, when a solution of the linearized KdV
equation
 itself is used in the rhs member, which part of it
is secular-producing, and which part is not. This is not too
difficult. Indeed, this solution is given by its expansion on the
basis of the squares $\Phi_k$ of the Jost functions related to the KdV equation \cite{kod78t,secular},
 and we have characterized the fact that a
source term is secular-producing or not by some criterion, that
involves the coefficients of this expansion and  their
$t$-dependency. It remains a last point to be studied: the
dependency of the
 higher order terms  with regard to
the higher order times. We shown  that it is governed by a
 linearized KdV Hierarchy  \cite{linkdv}. Finally, we have been able to justify
that the higher order terms are not  secular-producing, and  to
prove that the  formal expansion contains bounded terms only.

\section{Time scales}
The generalization of the above procedure to an arbitrary order $n\geqslant2$ yields
the  equation
\bq \frac{-1}{(-4)^nr_n}\partial_{\tau_n}\varphi_2=\partial_\xi{\cal L}^n\varphi_2,
\label{152}\eq
which governs the evolution of the main amplitude $\varphi_2$ with  regard to
the higher order time variable $\tau_n$.
 $\cal L$ is defined by the  above formula (\ref{hier7}). The scaling coefficient  $r_n$ is defined by $r_1=r$ and
the  recurrence formula
\bq
r_{n+1}=
\sum_{
\ba{c}\scriptstyle(\alpha_j)_{1\leqslant j\leqslant n-1},\,k\geqslant0\\
\scriptstyle
\left(\sum_{j=1}^{n-1}2j\alpha_j\right)+k=2n+3\ea}\hspace{-11mm}
\Xi((\alpha_j)_{1\leqslant j\leqslant n-1},k)
\prod_{j=1}^{n-1}(-r_j)^{\alpha_j}.
\label{153}\eq

The  sequence of time variables $\tau_1$, $\tau_2$, $\tau_3$,...
involved by the multiple time formalism are thus affected by the
sequence of scaling coefficients $r_1$,  $r_2$,  $r_3$,...~.
The equations of the KdV Hierarchy are 'universal', not specific to the physical situation considered.
 The time scaling coefficients contain thus most physical data about the time
evolution of the wave. Further, they are of interest regarding the
convergence of the asymptotic series. They are computed  using
recurrence formula (\ref{153}), together with the results
 of \cite{hierarchy} listed in the appendix. 
 The
first coefficients read as follows: \bq
\ba{rll}\displaystyle r_2=\frac{\gamma^3 V^9}{8(1 + \alpha)^2 m_t^4\mu }\biggl[&\displaystyle
  8  + (4\alpha - 13)\gamma  + (3\alpha+6)\gamma^2& \vspace{1mm}\\
        &\displaystyle +
           \alpha( 4\alpha-10)\gamma^3 + 4\alpha(1 -
            \alpha)\gamma^4&\biggr]
,\ea \eq
\bq\ba{rrl} \displaystyle
r_3=\frac{-\gamma^4V^{13}}{16(1 + \alpha)^3
m_t^6\mu^2}\biggl[\;\;40 +(32\alpha-88) \gamma +(8\alpha^2+ 12\alpha+67)\gamma^2& \vspace{1.5mm}\\
        \displaystyle
         +             (  68\alpha^2- 132\alpha - 20) \gamma^3 +
            (40\alpha^3\  -
            143\alpha^2+122\alpha+2) \gamma^4&  \vspace{2mm}\\
        \displaystyle-
\alpha( 32\alpha^2-70\alpha+ 32 )\gamma^5 +
           \alpha^2( 16\alpha^2- 56\alpha+6)\gamma^6&  \vspace{1.5mm}\\
        \displaystyle -
 8\alpha^2( 4\alpha^2 -
            8\alpha+1 )\gamma^7 +
            16\alpha^3(\alpha-1) \gamma^8&\biggr],
\ea\eq
in which
\bq
\gamma=1-\frac1{V^2}\quad,\qquad\mu=1+\alpha\gamma\quad,\qquad m_t=m\sin\theta.\label{2k10}\eq
 The expressions of the higher order coefficients can be
obtained in the same way, but are too complicated to be written
down here; numerical computation is more convenient.

\begin{figure}[hbt!]
\begin{center}
\includegraphics[width=10cm]{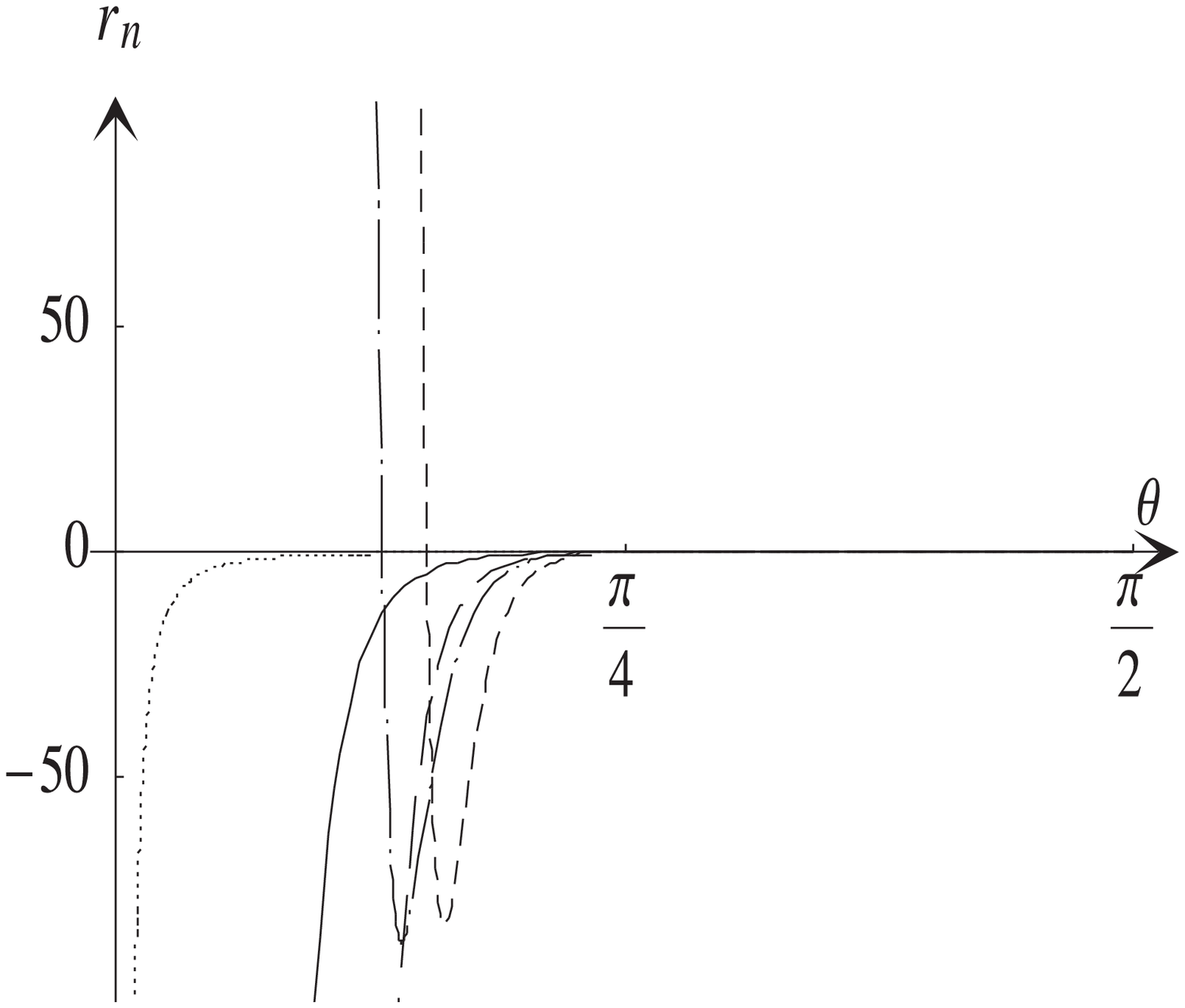}
\caption{\footnotesize  Plot of the five first time scaling
coefficients $r_1$, ..., $r_5$ against the angle $\theta$ between
the propagation direction and the exterior field. The rescaled
magnetic induction is $m=1$, and the parameter determining the
strength of the exterior field is $\alpha=0.5$~. Dotted line:
$r_1$, solid line: $r_2$, large dashing: $r_3$,  dashed-dotted
line: $r_4$, short dashing: $r_5$.} \label{fig1}
\end{center}
\end{figure}

The coefficients $r_1=r$, $r_2$,... up to $r_5$ are plotted on
figure \ref{fig1},
 against the value of the angle $\theta $
between the propagation direction and the external field, for a
given value of the parameter $\alpha$ that determines the
magnitude of this field. Notice the annulation and sign
change of the coefficients $r_4$ and $r_5$,
about 0.41 and 0.48 radians respectively. This  marks a 
change 
in the behaviour of the corresponding corrections.
 When $r_4$ is zero, $\varphi_2$ is constant with regard to $\tau_4$, thus
the $\rm 3^{rd}$ order correction is in fact valid at order 4,
regarding its time dependency.

It is seen that the $r_n$ take very small values when $\theta$ is
close to $\pi/2$, and very large values when $\theta$ is small. In
the limiting case where  the propagation direction  is orthogonal
to the external field ($\theta=\pi/2$), the velocity $V$ is 1,
thus $\gamma=0$,
and the coefficients $q$ and $r$ vanish, 
so that the KdV equation  (\ref{ferrikdv}) is replaced by \bq
\partial_\tau\varphi=0.\eq Thus $\varphi$ is constant with time at
fist order, which means {\it a priori} that the wave evolves much
slower than in the general case, at least for an order of
magnitude. Recall that this order of magnitude is determined by
the perturbative parameter $\e$, related to the wave amplitude and
typical length.
The wave propagates without deformation, to within 
a quantity of higher order in $\e$, up to times about $T/\e^3$,
instead of times about $T/\e$, as usual in the long wave
approximation. The higher order equations simplified by this
trivial time evolution of the main term, yield also an
approximation valid up to $T/\e^3$, for some finite $T$. Let us
precise the influence of the scaling coefficients on the time
validity range of the higher order approximations.
We denote by $L_0$ some typical length of the wave. The
dimensionless space variable is $\xi/{L_0}={\e (x-Vt)}/{L_0}$.
The reference length for $x$ is chosen with the order of magnitude of
$\e{L_0}$,
 in such a way that, as $x$
takes values as large as $1/\e$ with respect to this reference
length, $\xi$ is about $L_0$. $V$ is close  to 1, thus taking the
same value as a reference  time (recall that $t$ has already been
rescaled into  $ct$) is coherent with the asymptotic expansion.
The higher order time variables adapted to the expansion  are,
rather than the $\tau_n$, the variables $T_n$ of the KdV Hierarchy
(\ref{152}), written under its normalized form
\bq\partial_{T_n}v=\partial_\xi {\cal L}^nv\hspace{2cm} \mbox{($n$
integer)}\label{147}.\eq
 The
differential recurrence operator $\cal L$ is as in (\ref{hier7}),
with
 $v=\frac{q}{6r}\varphi_2$~.
The variable $T_n$ reads then:
\bq T_n=-(-4)^nr_n\tau_n=
-(-4)^nr_n\e^{2n+1}t\label{areu}\eq
 $T_n$ must be about 1 for
large values of $t$.
This necessitates a smaller value of $\e$ 
when the coefficient $r_n$, or rather $(-4)^n r_n$, is large.
For each $n$, 
$\e$ must be compared to the reference value $\e_n$ defined as
follows: $\e_n$ is the value of $\e$ in (\ref{areu}) such that,
 when $T_0=\e t$ is equal to $L_0$,
$\vert T_n\vert$ has the same value. It yields
\bq\e_n=\frac1{2r_n^{1/{2n}}}.\eq
The approximation involving the first $n$ time variables (for all
written terms) is valid for $t\leqslant T/\e^{2n+1}$, for some finite
$T$ with the order of magnitude of unity in the initial unit ($\e
L_0v$). Taking the scaling coefficients into account, the
approximation will rather be
 valid for $\vert T_n\vert\leqslant T$,  that is:
\bq\vert t\vert\leqslant \frac T{4^nr_n\e^{2n+1}}=\frac
T\e\left(\frac{\e_n}\e\right)^{2n} \label{p+1}\eq Notice that it
is in fact necessary that $\vert T_p\vert\leqslant T$ for all $p\leqslant
n$, what implies some conditions on the variations of $\e_p$ with
relation to $p$. According to (\ref{p+1}), when $\e_n$ takes large
values,  the
 propagation can be described  over a long distance
even if the  order $n$ is relatively low and the  value of the
perturbative parameter $\e$ close to 1.
The higher order time variables make sense only if $\e$ is smaller
than the $\e_n$, and  long time propagation can be described only
if the ratio $\e/\e_n$ is very small. These conditions will be
hardly fulfilled when $\e_n$ becomes small.

\begin{figure}[hbt!]
\begin{center}
\includegraphics[width=10cm]{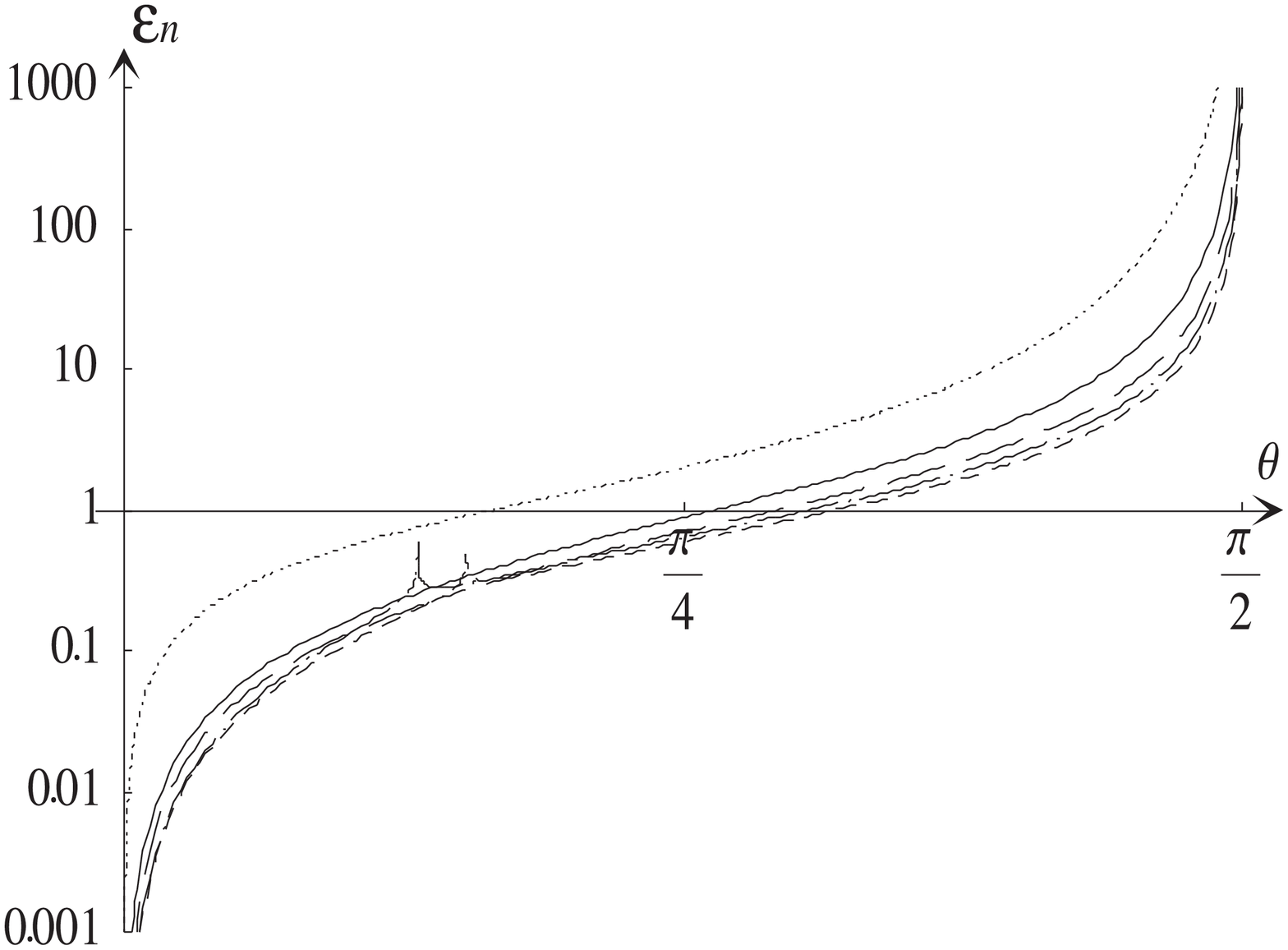}
\caption{\footnotesize  Logarithmic plot of the five first $\e_n$,
reference values for the perturbative parameter $\e$, built from
the time scaling coefficients $r_n$,
 against the angle
 $\theta$ between the propagation direction and the exterior field.
The value of the constants and the legend are the same as in
figure \ref{fig1}.} \label{fig2}
\end{center}
\end{figure}

\begin{figure}[hbt!]
\begin{center}
\includegraphics[width=10cm]{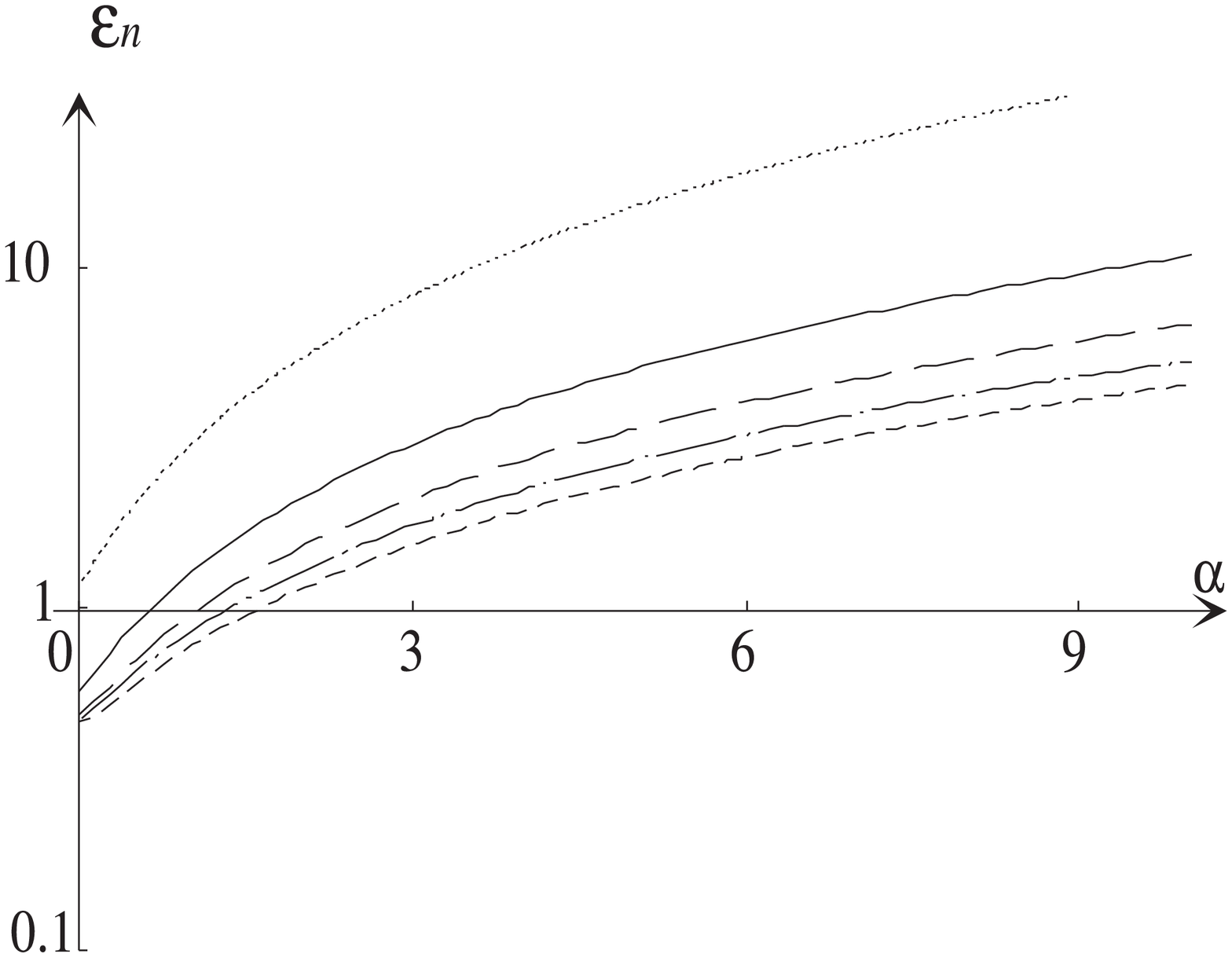}
\caption{\footnotesize  Logarithmic plot of  $\e_1$, ..., $\e_5$,
against the parameter $\alpha$ that determines the strength of the
exterior field. The rescaled magnetic induction is $m=1$, and the
angle
 between the propagation direction and the exterior field is  $\theta=\pi/4$~.
The legend is the same as in figure \ref{fig1}. } \label{fig3}
\end{center}
\end{figure}

The five first $\e_n$ are plotted on figure \ref{fig2},
 against
the angle $\theta$, and on figure \ref{fig3}, against the ratio
$\alpha$ that determines the magnitude of the external field. If
the extrapolation  of the few computed terms is valid, the
sequence $\e_n$ seems to be bounded with regard to $n$, although
its terms grow up as $\theta $ tends to $\pi/2$. Further, this bound
is not excessively small when $\theta$ is not smaller than $\rm
10^o$ or $\rm 15^o$. For smaller values of $\theta$, the $\e_n$
are so small that the KdV approximation can be valid only for
excessively low intensities, and the higher orders will never
appear.

 When $\theta$ approaches $\pi/2$,
the $\e_n$ become large. Then the approximation yielded by the KdV equations
is valid during a very long time. 
The pulse behaviour will be correctly described by them even if  the small
perturbative parameter  $\e$ takes values rather close to 1. At
the limit $\theta=\pi/2$, the modulation described by the KdV
equation itself arises only at a very slow rate.
A typical dependency of the $\e_n$ with regard to the strength of the external field
is shown on figure \ref{fig3}. The $\e_n$ grow slowly with $\alpha$.
Thus a strong external field enhances the validity of the KdV approximation, and
increases duration along which it can be expected to describe the physics. However this effect is  much weaker than the
dependency with regard to the direction of the external field and the angle
$\theta$.

\section{The soliton of the hierarchy}
The time scaling coefficients studied in the previous paragraph have given an insight into the convergence of the
 perturbative expansion as an asymptotic behaviour
for small values of the perturbative parameter $\e$, for a fixed number $n$  of corrective terms.
We are also able to get some insight into the convergence of the series when $n$ tends to infinity and $\e$ is fixed,
through the computation of the one soliton solution of the complete KdV hierarchy.
As mentioned above, all equation of the KdV hierarchy are
compatible together, in the sense that for a given initial data, a
function $v(X,T_1,T_2,T_3,\cdots)$ satisfying  equation
(\ref{hier6}) for any value of $n$ can be found. This
solution can be found using  the inverse scattering transform
(IST) method, at least in principle. Indeed, all equations of the
hierarchy are completely integrable by means of the IST method.
Furthermore, they can all
 be described in the IST
formalism using the same spectral problem (\cite{dodd}, p. 96), which
ensures their compatibility. The scattering
data $\left(R_+(k),D_{+,j},k_j\right) $ (see \cite{dodd}, p. 141 {\it sq.}, for the precise definition of these quantities)
 are defined in the same way for all
 equations, only their time evolution differ for each time
variable $T_n$. These time evolution is
given by (\cite{dodd}, p. 149)
\begin{eqnarray}
R_+(k,T_n)&=&R_+(k,0)e^{\Omega_n(k)T_n},\label{sol1}\\
D_{+,j}(T_n)&=&D_{+,j}(0)e^{\Omega_{n,j}T_n},\label{sol2}\\
k_j(T_n)&=&k_j(0).\label{sol3}
\end{eqnarray}
The  index $n$ refers to the $n^{\rm th}$ equation of the hierarchy.
The evolution factors are $\Omega_{n,j}=\Omega_n(k_j)$, and  $\Omega_n(k)=-i\omega_j(2k)$,
where $\omega_j(k)$ is the dispersion relation of the  $n^{\rm th}$ equation of the hierarchy linearized.
Its seen from relation (\ref{sol3}) that the discrete spectrum $(k_j)$ is constant with regard to any of the time variables
$T_n$. Therefore the number
of solitons and their characteristics are not modified by the higher order time evolution.
The evolution of the spectral data with regard to all the higher order time variables can
then be written as a single exponential factor for each spectral component,
\bq
R_+(k,T_1,T_2,\cdots)=R_+(k,0,0,\cdots)\exp\left(\sum_{n=1}^\infty\Omega_n(k)T_n\right).\label{sol4}\eq
From the expression (\ref{hier6}-\ref{hier7}) of the equations of the hierarchy, we find that
\bq\omega_n(k)=\frac{-k^{2n+1}}{4^n}\label{sol5}.\eq
For a  value of the spectral parameter  $k$ belonging to the discrete spectrum, $k=k_j=i\kappa_j$ with $\kappa_j$ real,
we get
\bq\Omega_{n,j}=2(-1)^{n+1}\kappa_j^{2n+1}.\label{sol4.5}\eq
Using the definition (\ref{areu}) of the time variable $T_j$, we get the following expression
 of the complete time evolution factor:
 \bq
\sum_{n=1}^\infty\Omega_{n,j}T_n=\Omega_j t,\quad \mbox{with}\qquad
\Omega_j=\sum_{n=1}^\infty\left(2\e\kappa_j\right)^{2n+1}r_n.\label{sol6}\eq
Obviously  formula (\ref{sol6}) is valid only if  the power series converges. Notice that the
coefficients of the latter are the time scaling coefficients $r_n$.
For a one-soliton solution, the above formulas show that the introduction of a sequence of higher order
time variables and of all equations of the KdV hierarchy yield nothing but a
renormalization of the soliton speed. This result can be also found by direct computation as follows.
By definition, the one-soliton solution  propagates without deformation, at least with regard to the
first time variable $T_1$.
It can thus be written under the form $v=v(X+\lambda T_1)$.  Then using the KdV equation, {\it i.e.}
equation (\ref{hier6}) with $n=1$, we see that $v$ is an
eigenvector of the recurrence operator $\cal L$ defined by (\ref{hier7}), with the eigenvalue $\lambda$.
We deduce easily the $T_n$-dependency of $v$, it is given by  $v=v(X+\lambda^n T_n)$. 
We find this way the expression of the one-soliton solution of the complete hierarchy:
\bq
v=2b^2\sech^2\,b\left(X+\sum_{n=1}^\infty\left(-b^2\right)^n T_n\right),\label{sol7}
\eq
using the normalized variables.
In the case of magnetic solitons, it can be
written using the physical variables as
\bq \vec H_w=\frac{12 r}q\vec h_1\beta^2\sech^2\,\beta(x-{\cal V}t),\label{sol8}\eq
where
\bq {\cal V}=V+\sum_{n=1}^\infty 4^n\beta^{2n}r_n.\label{sol9}
\eq
$V$ is the velocity given by (\ref{vit}), $\vec h_1$ the polarization vector given by (\ref{achun}).
The wave magnetic field $\vec H_w$ is related to the previously defined field components
 through
\bq \vec  H=\vec H_0+\e^2\vec H_2+\cdots\simeq\vec H_0+\vec H_w.\label{sol10}\eq
The dimensional soliton parameter $\beta$ is related to the normalized soliton parameter $b$ through
$\beta=\e b$.
Computation of the one-soliton from the IST formalism allows to identify the soliton parameter $b$ to the
single discrete eigenvalue $\kappa_1$. This way we check that the  relative soliton velocity
$\left({\cal V}-V\right)$
given by (\ref{sol9}) is equal to $\Omega_1/(2\beta)$, using the expression (\ref{sol6}) of the evolution factor $\Omega_1$.

The soliton speed is thus given by a power series of the soliton parameter $\beta$, whose coefficients
are essentially the time scaling coefficients $(r_n)_{n\geqslant1}$.
Obviously if this series diverges so does the whole perturbative scheme. Reciprocally, the convergence
of the series defining the velocity should favour that  of the perturbative scheme,
although the latter is by no means proven.
Writing the power series which defines $\cal V$  as
\bq {\cal V}=V+\sum_{n=1}^\infty\left(\frac\beta{\e_n}\right)^{2n},\eq
we
see that it converges when
\bq\beta<\beta_M=\liminf_{n\longrightarrow\infty}\e_n\eq
and diverges for larger values of the soliton parameter $\beta$.
Therefore the limit of the sequence $\e_n$ for  large $n$ gives us a maximal value $\beta_M$
of the soliton parameter $\beta$, above which
we know that the perturbative scheme does not converge when increasing the number of terms.
Physically, this lack of convergence  means that the
KdV soliton will be destroyed by some  effects which cannot be taken into account using the perturbative approach.

For values of the soliton parameter $\beta$ below the limit $\beta_M$, we get a renormalized soliton speed,
{\it a priori} valid  for an infinite propagation time.
The boundness of all terms in the perturbative scheme proves that for a given propagation time
and a given number of terms, this soliton gives
a good approximation of the real impulsion for small enough values of $\e$, {\it i.e.} of $\beta$.
 We can reasonably conjecture
that small enough can be understood here as less than the limiting value $\beta_M$ of $\beta$.
Physically it means that magnetic KdV solitons with parameter smaller than $\beta_M$ should conserve their properties
during a long propagation time.

\begin{figure}[hbt!]
\begin{center}
\includegraphics[width=10cm]{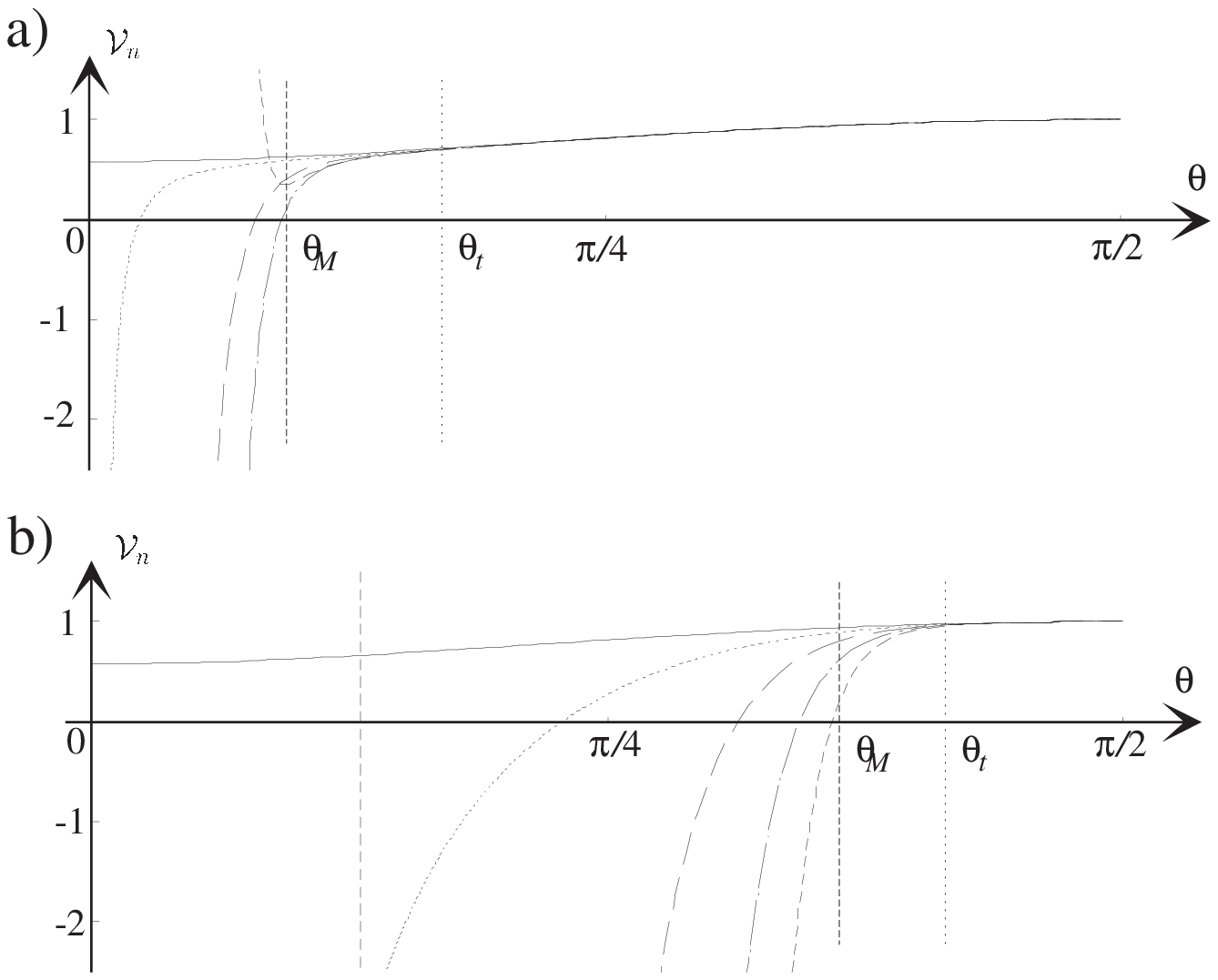}
\caption{\footnotesize  Plot of the five first approximate values ${\cal V}_n=V+\sum_{p=1}^n\left(\beta/{\e_n}\right)^{2n}$
 of the soliton velocity ${\cal V}$, against the angle $\theta$ between
the propagation direction and the exterior field. The rescaled
magnetic induction is $m=1$, and the parameter determining the
strength of the exterior field is $\alpha=0.5$~.
Solid line: $n=0$, dotted line: $n=1$, large dashing: $n=2$,  dashed-dotted
line: $n=3$, short dashing: $n=4$. For a soliton parameter $\beta=0.1$ (a), $\beta=1.5$ (b). } \label{fig4}
\end{center}
\end{figure}

According to figures \ref{fig2} and \ref{fig3},
 the $\e_n$, thus also their limit $\beta_M$, depend on the physical parameters, and specially on the
angle $\theta$ between the propagation direction and the applied field.
 An example of computation showing the convergence of the velocity series is drawn on figure \ref{fig4}
as a function of this angle. It is seen that, for $\theta$ close to $\pi/2$, the first approximation (KdV)
gives almost the exact speed, while for small angles the series diverges.
We denote by $\theta_M$ the value of $\theta$ for which $\beta_M$ is equal to the fixed value of $\beta$.
When $\theta< \theta_M$,  the series does not converge, and the whole perturbative  approach is not valid.
 To compute $\theta_M$ for the figure \ref{fig4}, we have approximated $\beta_M$ by $\e_5$.
When $\theta> \theta_M$, if we consider only the soliton speed, the KdV approximation
will correctly describe the wave evolution. More precisely, the KdV equation itself
will give an acceptable description above some value  $\theta_t$   of the angle $\theta$,
while this first order approximation  needs
to be corrected by higher order terms below $\theta_t$ (notice that the threshold value $\theta_M$ is precisely defined,
 while $\theta_t$ is only an order of magnitude  depending on the
accuracy required).
It is reasonable to think that the same kind of conclusion holds in a more general situation,
 involving several solitons and radiation.

\section{Conclusion}
The multiple time formalism has been applied to the study of the propagation of
KdV solitons in ferromagnetic media. According to this formalism, the
dependency of the
higher order terms with respect to the first order  time variable is given by linearized KdV equations,
while the dependency of the main term with regard to the higher order time variables is
governed by all equations of the KdV Hierarchy.
 The latter  are determined by the
requirement that the linear terms in the rhs of the linearized
KdV equation vanish. This yields  scaling coefficients for  the higher order time
variables of the KdV Hierarchy, which contain most physical information
concerning the wave evolution. Explicit computation of these scaling
coefficients shows in particular that the approximation yielded by
the KdV model gives a good account for the physical behavior of
the wave during long propagation times when the angle between the
propagation direction and the external field is large enough.
 The time during which the pulse is correctly described by
the KdV equation falls to zero when they are parallel.
Mathematically, the perturbative parameter $\e$ is infinitely small, while it takes
a finite value in a physical situation. The approximation is valid only if this finite value is small
enough. The corresponding range of the perturbative parameter $\e$ is usually determined in a rather empirical way.
The present study gives some theoretical
insight into this question, through a physical interpretation of the time scaling coefficients.

The one-soliton solution of the complete KdV hierarchy
has been written down as a function of the physical parameters.
The soliton velocity writes as a power series of the soliton parameter,
involving the sequence of the time scaling coefficients.
We get a maximum value of the soliton parameter, above which the perturbative series diverges.
Then the KdV approximation, even with corrective terms, does not describe correctly the physics.
If the soliton parameter is below the threshold, the long-distance effect of the higher order corrections is only a
modification of the soliton speed, and the physical system  behaves qualitatively as the KdV model.
It has been observed that the validity domain of KdV-type asymptotics
is often much larger than predicted by the mathematical analysis. The above conclusions can partially explain
this observation: the KdV-type behaviour is qualitatively correct in
the whole  validity domain of the infinite  KdV hierarchy expansion,
which is expected to be much larger.

\newpage
\appendix{\LARGE\bf Appendix}
\vspace{1cm}

We list in this appendix the formulas needed for the computation of the time scaling coefficients
$r_n$. These formulas are proven in \cite{hierarchy}.
$r_n$ is given by equation (\ref{153}) with
\bq\ba{rl}\displaystyle\hspace{-5mm}
\Xi((\alpha_j)_{j\geqslant1},k)
=&\displaystyle\hspace{-2.5mm}
\frac{-1}\Lambda\biggl[ V\vec m\cdot\vec m\;((\alpha_j)_{j\geqslant1},k-1)
\vspace{1.5mm}\\&\displaystyle\hspace{-13mm}
-\sum_{i\geqslant1}
\vec m\cdot\vec m\;((\alpha_j-\delta_{i,j})_{j\geqslant1},k-1)\biggr],\ea\label{89}\eq
where
\bq\vec m=\left(\ba{c}m_x\\m_t\\0\ea\right).\eq
$\vec m\;((\alpha_j),k)$ is deduced according to
\bq\tilde u\;((\alpha_j)_{j\geqslant1},k,l)=\left(\ba{c}
\vec e\;((\alpha_j)_{j\geqslant1},k,l)\\
\vec h\;((\alpha_j)_{j\geqslant1},k,l)\\
\vec m\;((\alpha_j)_{j\geqslant1},k,l)\ea\right),\eq
from the following recurrence formulas.
For all
 $l\geqslant1$,
\bq \tilde u\;((0),0,l)=\tilde u_1.
\label{82}\eq
For all $k$ and $l\geqslant1$,
\bq
 \tilde u\;((0),k,l)=S(V\vec m\;((0),k-1,l)).
\label{83}\eq
 For all $(\alpha_j)_{j\geqslant1}\neq(0)$
and $l\geqslant1$,
\bq\ba{l}\displaystyle
 \tilde u\;((\alpha_j)_{j\geqslant1},0,l)=\vspace{1mm}\\
\hspace{1.5cm}\displaystyle
\sum_{i\geqslant1}\Phi(\tilde u\;((\alpha_j-\delta_{i,j})_{j\geqslant1},0,l))\ea
\label{84}.\eq
For all $(\alpha_j)_{j\geqslant1}\neq(0)$, $k$, $l\geqslant0$,
\bq\ba{l}\hspace{-3mm}
 \tilde u\;((\alpha_j)_{j\geqslant1},k,l)=\vspace{1.5mm}\\
\displaystyle\hspace{0.9cm}
S\,(V\vec m\;((\alpha_j)_{j\geqslant1},k-1,l))\vspace{1.5mm}\\
\displaystyle\hspace{0.9cm}
+\sum_{i\geqslant1}
\biggl[\Phi(\tilde u\;((\alpha_j-\delta_{i,j})_{j\geqslant1},0,l))\\
\displaystyle\hspace{1.5cm}
- S\,(\vec m\;((\alpha_j-\delta_{i,j})_{j\geqslant1},k-1,l))\biggr].\ea
\label{85}\eq
In (\ref{82}-\ref{85}), $\delta_{i,j}$ is the Kronecker symbol, $V$ is given by (\ref{vit}). $\Phi$ is defined by
\bq\Phi(u)=S(\alpha\vec m\wedge\vec m_I(u))+u_I(u)\label{54},\eq
where $S$
is the $9\times 3$ matrix
\bq S=TL^{-1}\hspace{1cm}\mbox{with}\hspace{1cm}T=\left(\ba{c}-\frac1VR_x\\
I\\-\Gamma\ea\right)\label{39}.\eq
$I$ is the three-dimensional unity matrix,
and
\bq L^{-1}=\frac1{\mu m_xm_t}\left(\ba{ccc}0&0&0\\0&0&m_t\\
m_x&0&0\ea\right)\label{35}.\eq $u_I$ is the linear operator in
${\mathbb R}^9$ defined by \bq u_I(\left(\ba{c}\vec E\\\vec
H\\\vec M\ea\right))
=\left(\ba{c}\frac1V\vec E\\0\\
\vec m_I(\left(\ba{c}\vec E\\\vec H\\\vec M\ea\right))\ea\right),\label{40}\eq
with
\bq\vec m_I(\left(\ba{c}\vec E\\\vec H\\\vec M\ea\right))=
\frac1V(\vec H+\vec M)+\frac1{V^2}R_x\vec E\label{26}\eq
$R_x$ is the $3\times 3$ matrix
\bq R_x=\left(\ba{ccc}0&0&0\\0&0&-1\\0&1&0\ea\right)\eq

The first term of the sequence $ \tilde u\;((\alpha_j)_{j\geqslant1},k,l)$ is given by
$\tilde u_1=T\vec h_1$, where $\vec h_1$ is the polarization vector defined by (\ref{achun}),
which also reads
\bq\vec h_1=\left(\ba{c}\mu m_x\\(1+\alpha)m_t\\0\ea\right).\eq
The quantity $\Lambda$ in (\ref{89}) is given by
\bq \Lambda=V\vec m\cdot\vec\Phi_m(\tilde u_1),\label{60}\eq
where $\Phi_m$ is the $m$-component of $\Phi$ defined by (\ref{54}),
according to
\bq\Phi=\left(\ba{c}\vec \Phi_e\\\vec \Phi_h\\\vec \Phi_m\ea\right).\eq
We use the shortcuts
\bq\Gamma=\left(\ba{ccc}1&0&0\\0&\gamma&0\\0&0&\gamma\ea\right)\quad,\qquad m_x=m\cos\theta,\eq
and $\gamma$, $\mu$, $m_t$ given by (\ref{2k10}).
The expression (\ref{vit}) of the velocity yields the relation
\bq\mu m_x^2+\gamma(1+\alpha)m_t^2=0\label{45},\eq
which is useful to simplify the expressions.

\end{document}